\documentclass[mathleft
]{an}
\usepackage{graphicx}
\usepackage{times}
\usepackage{enumerate}
\usepackage{amsmath}
\newcommand\leftidx[3]{%
  {\vphantom{#2}}#1#2#3%
}
\begin{document}

\Pagespan{43}{48}
\Yearpublication{2014}%
\Yearsubmission{2014}%
\Month{}%
\Volume{}%
\Issue{}%
 \DOI{}%

\title{Mini-Review on Mini-Black Holes from the Mini-Big Bang
    }

\author{M. Bleicher\inst{1,2}
\and  P. Nicolini\inst{1,2} \fnmsep\thanks{Corresponding author:
  \email{nicolini@th.physik.uni-frankfurt.de} 
  }}
\titlerunning{Mini-Review on Mini-Black Holes from the Mini-Big Bang}
\authorrunning{M. Bleicher \& P. Nicolini}
\institute{Frankfurt Institute for Advanced Studies, Ruth-Moufang-Str. 1, D-60438 Frankfurt am Main, Germany
\and
 Institut f\"{u}r Theoretische Physik, Johann Wolfgang Goethe-Universit\"{a}t, 
Frankfurt am Main, Germany }

\received{} \accepted{} \publonline{later}

\keywords{Black Hole Physics, Extra-Dimensions, Terascale Quantum
Gravity}

\abstract{%
  We review the basic ideas about man-made quantum mechanical black holes. We start by an overview of the proposed attempts to circumvent the hierarchy problem. We study the phenomenological implications of a strong gravity regime at the terascale and we focus on the issue of microscopic black holes. We provide the experimental bounds on relevant quantities as they emerge from major ongoing experiments.  The experimental results  exclude the production of black holes in collisions up to 8 TeV. We provide some possible explanations of such negative results in view of forthcoming investigations.}

\maketitle

\section{Introduction}
Gravity is certainly the most commonly known fundamental interaction. In every moment  we experience our body weight and we are familiar with the fall of objects. No other fundamental interaction affects everyday life so widely and continuously as gravity. Also the absence of gravity does not come as a surprise. Space explorations made popular this concept, we often  call ``weightlessness''.
However, despite the popularity, gravity is probably the  less-understood fundamental interaction. Such incomplete understanding is already evident from the more elementary concepts. For instance the gravitational constant, $G = 6.67384(80) \times 10^{-11} \ {\rm N}\, {\rm (m/kg)^2}$ (\cite{MTN11}), is currently known with an accuracy slightly better than that of the original Cavendish experiment in 1798 (\cite{Gil97}). The difficulties in performing accurate measurements are explained by the fact that gravity is extraordinarily weak and becomes sizable only if huge bodies are taken into account. Accordingly measurements of gravity become even more complicated  at short scales.  At typical scales of particle physics gravity is simply negligible. Even for the heaviest elementary particles (\textit{e.g.} W bosons, top quarks) electromagnetic interactions exceed   gravitational forces for something like $32$ orders of magnitude.

\begin{figure}
\includegraphics[width=0.475\textwidth]{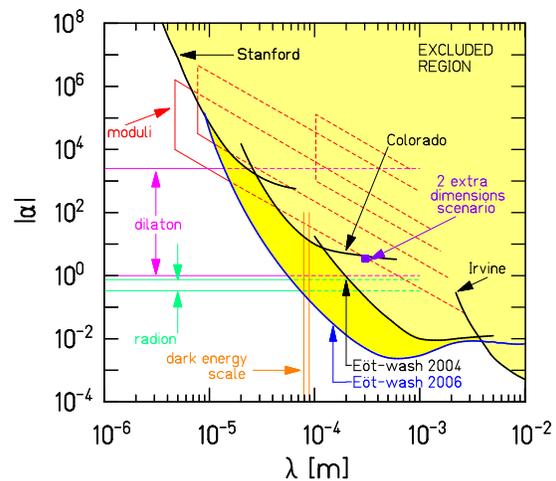}
\caption{Deviations from the inverse square law (\cite{LCC03}).}
\label{figure1}
\end{figure}

The reason why gravity is so weak is an open issue that is usually labeled as ``hierarchy problem''.  A possible way to overcome this difficulty is based on the idea that gravity, even if weak at macroscopic scales, might be strong at some minuscule scale. Such a conjecture is admissible because we cannot experimentally measure gravity at scales below some tens of microns (\cite{LCC03}; \cite{HKH04}; \cite{AHH07}; \cite{AGH09}). As a result one can contrive a mechanism in order to increase the strength of gravity, by postulating a different spacetime structure at microscopic scales. This opportunity is offered by the so called ``extra-dimensions'', \textit{i.e.}, additional spatial dimensions that have to be small enough to  be conventionally unobserved. Extra-dimensions are the natural theoretical framework for Superstring theory, Supergravity, M-theory  but have a long history dated back to 1921, the year in which the Kaluza-Klein (KK) theory was presented. To solve the hierarchy problem, however, the typical size, $R$, of each additional dimension cannot be arbitrarily small, \textit{ e.g.} $R\sim 1/m_\mathrm{P}$, where $m_\mathrm{P}\sim 10^{16}$ TeV is the Planck mass. Rather we require  $R$ to be large enough in order to ``lower down'' the Planck mass to a new fundamental mass, $M_\mathrm{F}$, which has to be above the electroweak scale, $\Lambda_\mathrm{EW}\sim 200$ GeV, but at the reach of (or at least not too far from)  current particle accelerator energies. Following this reasoning, one would like to place $M_\mathrm{F}$ at the terascale, \textit{i.e.}, $M_\mathrm{F}\sim 1-10$ TeV.
In addition, to reproduce known results from  particle physics,
one requires that only gravity can probe extra-dimensions, while conventional Standard Model fields must be constrained on a $(3+1)$-dimensional sub-manifold, called \textit{brane}\footnote{For sake of completeness, we recall that models with \textit{universal} extra-dimensions allow the propagation of all fields in the full higher-dimensional spacetime (\cite{ACD01}).}.

\section{Extra-dimensions: why and how}

Along the above line of reasoning, in the late 1990's\footnote{Antoniadis (1990) first proposed the idea almost a decade earlier.} Antoniadis et al. (1998),  Arkani-Hamed, Dimopoulos \&  Dvali (1998a, 1998b) presented the  model of \textit{large} extra-dimensions (also known as ADD model) as a viable solution to the hierarchy problem. The basic ingredient of such an approach is a $(4+d)$-dimensional \textit{bulk} spacetime, ${\cal M}^{(4+d)}$, that can be factorized as ${\cal M}^\mathrm{(4+d)}={\cal M}^{(4)}\times T^{(d)}$, where  ${\cal M}^{(4)}$ is the brane and $T^{(d)}$ is a $d$-dimensional torus with radii of size $R$. From the higher-dimensional Einstein-Hilbert action, one can obtain a dimensionally reduced action by performing a KK expansion of the graviton field. Since the zero mode is independent of extra-dimensional variables, one can integrate the $d$ additional dimension to obtain  (\cite{Shi10})
\begin{eqnarray}
&&S_{d+4}\sim M_\mathrm{F}^{d+2}\int {\cal R}\sqrt{-g}\ \mathrm{d}^4 x\ \mathrm{d}^d y\\
&&\to  \underbrace{M_\mathrm{F}^{d+2}R^d \int \leftidx{^{(4)}}{{\cal R}}{}\sqrt{-\leftidx{^{(4)}}{g}{}}\ \mathrm{d}^4 x}_{\mathrm{effective\ brane\ action}} +\underbrace{\sum_{k>0}\left(\dots\right)}_{\mathrm{KK\ excitations}}.\nonumber
\end{eqnarray}
The effective $4$-dimensional action has to match the standard gravitational action for distances larger than $R$. This matching sets the fundamental scale as $M_\mathrm{F}\sim \left( m_\mathrm{P}^2/R^d\right)^{1/(d+2)}$. Conversely for distances smaller than $R$ one can find compelling deviations to Newton's law
\begin{equation}
V(r)\sim \frac{1}{M_\mathrm{F}^{d+2}}\frac{m_1m_2}{r^{d+1}},\quad r< R,
\end{equation}
where $m_1$ and $m_2$ are two test masses at distance $r$.
By setting $M_\mathrm{F}\sim 1$ TeV, one obtains the size of extra-dimensions as $R\sim 10^{\frac{32}{d}-19}$ m. The cases $d=1,\ 2$ are immediately ruled out since they would require too large values for $R$. On the other hand for $d>2$, sub-millimeter measurements of the inverse square law become rather ineffective in setting parameter bounds (see Fig. \ref{figure1}). We remark here that the emission of graviton KK exitations can be constrained with astrophysical tests like the cooling of hot stars \textit{e.g.} the supernova SN 1987A. Resulting bounds on extra-dimension turn to be more stringent. The cases $d=1-4$ are ruled out but for $d>4$ the limits are rather loose (\cite{FrG08}).

An alternative proposal\footnote{For sake of brevity we cannot recall here all alternative mechanisms to tackle the hierarchy problem, which include, among the others, split fermion models (\cite{ArS00}; \cite{AGS00}; \cite{MiS00}) and un-particle enhancement models (\cite{Mur08}).} to the ADD model is offered by the \textit{warped} extra-dimensions. Following the lines of Randall \& Sundrum (RS) (1999a)\footnote{Gogberashvili (1999, 2000, 2002) proposed an analogue set up in terms of ``thin shells''.}, one postulates the existence of just one extra-dimension to obtain a $5$-dimensional warped anti-deSitter geometry
\begin{equation}
\mathrm{d}s^2=e^{-2\kappa R|\varphi|}\eta_{\mu\nu}\,\mathrm{d}x^\mu\, \mathrm{d}x^\nu +R^2\, \mathrm{d}\varphi^2
\end{equation}
where $\kappa$ is a scale $\sim{m}_\mathrm{P}$,  $x^\mu$ are the conventional $4$-dimensional coordinates, while $0\leq \varphi \leq \pi$ is the coordinate of the extra-dimension whose size is $\pi R$. At the boundary of the above $5$-dimensional spacetime, there are two branes\footnote{Randall \& Sundrum (1999b) later proposed  a single brane model.} called Planck (or hidden) brane and TeV (or visibile) brane, for $\varphi=0,\pi$ respectively.
The fundamental mass turns out to be of the order of the Planck masss, $M_\mathrm{F}\sim m_\mathrm{P}$. However this is not a source of concern because any mass parameter $m_0$  will correspond on the visible brane to a physical mass suppressed by the warp factor, $m=e^{-\kappa R\pi}m_0$.  Contrary to the ADD model, the exponential factor lets us have a large hierarchy of scales without extremely large values of $R$. If $R$ is just few tens of Planck lengths $l_\mathrm{P}\equiv 1/m_\mathrm{P}\sim 10^{-35}$ m, the fundamental scale will be lowered down to the TeV regime on the visible brane. In addition, there are no KK light modes (but only TeV KK modes) and constraints from supernova cooling do not apply.

In conclusion,  extra-dimensional models are more effectively tested if one probes small distances by means of high energy physics experiments (\cite{Fen03}).

\section{Black holes at colliders}

Black holes (BHs) are conventionally known as spacetime regions of no escape, delimited by what is technically known as an event horizon. BHs result from solutions of Einstein's field equations and their actual existence is corroborated by the astronomical observations of several candidate BH objects (\cite{CMS99}). The size of BHs is set by their gravitational radius $r_\mathrm{g}\sim GM_\mathrm{BH}/c^2$.
Depending on their mass, we usually distinguish stellar BHs,  intermediate mass BHs and supermassive BHs.  The total mass range is rather wide, \textit{i.e.}, $M_\mathrm{BH}\sim 10-10^{10}\ M_\odot$ where $M_\odot\sim 10^{30}$ kg is the solar mass. Accordingly also their formation mechanism is expected to differ drastically. If one considers even smaller masses and radii, \textit{i.e.},  $M_\mathrm{BH}<M_\mathrm{Moon}\sim 10^{23}$ kg  and $r_\mathrm{g}<0.1$ mm, tiny BHs might have formed due to extreme local matter density fluctuations in the primordial Universe (\cite{CaH74}). Alternatively such primordial BHs might have been pair produced from the quantum mechanical decay of the deSitter early Universe (\cite{MaR95}; \cite{BoH96}).

The extreme conditions of such scenarios occurring in the  early Universe provide some initial clues about
%
%
how difficult is to artificially produce a microscopic black hole in a particle collision. Following the  ``hoop conjecture'' (\cite{Tho72}), one finds that the collision energy should exceed $m_\mathrm{P}$ to produce a BH, an occurrence that is far above accessible experimental energies. Conversely, if gravity is strong at  shortest scales, gravitational collapses become possible at energies at the reach of current accelerators, \textit{i.e.},  $\sim M_\mathrm{F}$ (\cite{BaF99}; \cite{BHH02}). One may wonder, however, what extra-dimension set up has to be used for a consistent BH description. Technically the ADD model has the advantage of  BH geometries that do not suffer of significant tidal distortions as in  warped extra-dimension  models (\cite{Maa04}).   This is a direct consequence of the largeness of extra-dimensions. The size of BHs in the ADD model, $r_\mathrm{g}\sim 10^{-19}\ll R$ permits to neglect  manifold boundary effects.

The first papers concerning the concrete possibility of producing BHs at the LHC are due to
Argyres, Dimopoulos \& March-Russell (1998), Banks \& Fischler (1999), Dimopoulos \& Landsberg
 (2001) and Giddings and Thomas  (2002). Nowadays there exists a huge number of papers, reviews and books about several aspects of the topic, that we cannot address in the present contribution. For an ``incomplete'' list of suggested readings see (\cite{Lan02}; \cite{Cav03}; \cite{Kan04}; \cite{Hos04}; \cite{CaS06}; \cite{Ble07}; \cite{Win07}; \cite{Nic08};  \cite{BlN10}; \cite{Cal10}; \cite{Par12}; \cite{KaW14}).

 From the static, hyper-spherical BH geometry (\cite{Tan63}), one has
 \begin{equation}
r_\mathrm{g}^{d+1}=\left(\frac{1}{M_\mathrm{F}\sqrt{\pi}}\right)^{d+1}\left(\frac{M_\mathrm{BH}}{M_\mathrm{F}}\right)\left[\frac{8\Gamma\left(\frac{d+3}{2}\right)}{d+2}\right].
\end{equation}
One can see that, for $M_\mathrm{BH}\approx M_\mathrm{F}$,  the gravitational radius $r_\mathrm{g}\sim 1/M_\mathrm{F}$ weakly depends on the number of dimensions. In addition such profile of $r_\mathrm{g}$ confirms that large extra-dimensions are a crucial ingredient for enhancing the BH production rate. The decrease of the gravitational scale from $m_\mathrm{P}$ to $M_\mathrm{F}$ implies an increase of the BH cross section $\sigma(XX\to \mathrm{BH})\sim\pi r^2_\mathrm{g}$. Despite the intense discussion in the literature about a variety of proposed modifications (\cite{Cav03}), the above black disk profile is widely accepted, at least for collision energies far above $M_\mathrm{F}$ (\cite{MNS12}). Accordingly the estimated cross section value is $\sigma\sim 400$ pb.
   Given the LHC design luminosity $L\sim 10^{38}$ m$^{-2}$ s$^{-1}$,  about a hundred BHs per second would form in particle detectors, or equivalently a billion BHs per year (\cite{HHB02}; \cite{BHH02}).

The life of a BH in a particle detector is a problem of formidable complication.
Mini BHs behave rather differently from their macroscopic counterparts. At microscopic scales quantum mechanical effects cannot be neglected. As shown by Hawking (1975), this opens the possibility of tunneling particles through the even horizon, which acts like any other potential barrier. The resulting effect, known as BH evaporation, is a thermal emission of particles at a temperature $T_\mathrm{H}\sim 1/r_\mathrm{g}$, with a consequent BH decay. In general, the description of these issues would require a quantum theory of gravity, since the spacetime itself is subject to relevant quantum modifications at these scales. For instance string theory encodes modifications related to the idea of non-commutative geometry (\cite{SeW99}) or the generalized uncertainty principle (\cite{Ven86}). Accordingly one can model these characters by an effective implementation  in BH spacetimes (\cite{NSS06}; \cite{IMN13}).  However, in the trans-Planckian regime, \textit{i.e.}, for $M_\mathrm{BH}\gg M_\mathrm{F}$, the semi-classical approximation of quantum gravity can still be used for drawing scenarios of BH evaporation.

For pedagogical purposes, one can distinguish four phases of the  life of a mini-BH after its formation
\begin{enumerate}[i)]
\item Balding phase, during which the gauge field hair is shed
and the asymmetries removed by gravitational radiation;
\item Spin-down phase, during which the black hole
evaporates through Hawking and Unruh-Starobinskii (\cite{Sta73}; \cite{Unr74}) radiation, losing
mostly angular momentum and mass;
\item Schwarzschild phase, during which the black hole keeps evaporating
via Hawking radiation, but now in a spherical manner;
\item Planck phase, during which $M_\mathrm{BH}\sim T_\mathrm{H}\sim M_\mathrm{F}$ and quantum gravity effect cannot be neglected.
\end{enumerate}
The destiny of an evaporating BH is uncertain. However there are two prevailing scenarios. The first possibility is that the BH concludes its life by a non-thermal emission of particles (\cite{EHM00}; \cite{GiT02}; Calmet, Fragkakis \& Gausmann 2012). Alternatively, the BH might undergo a cooling down phase towards a zero temperature remnant configuration (\cite{HBH03}). The latter scenario seems to be a model independent character common to several spacetime models, derived or inspired by quantum gravity considerations (\cite{BoR06}; \cite{Mod06}; Nicolini et al. 2006; \cite{MMN11}; \cite{Nic12};  Isi et al. 2013). If this were the case, the remnant formation would affect the emission spectra:\footnote{Remnants can form also in the absence of a cooling down, like in the case of hot Planckian remnants (\cite{APS01}) or following dimensional reduction mechanisms (\cite{Mur12}; \cite{MuN13})} any cooling down phase markedly implies an emission of softer particles mostly on the brane (\cite{Gin10}; \cite{NiW11}).

The properties of emission spectra are included in the more general framework of the signatures of BH production. To this purpose we recall that the Hawking emission is not expected to be directly observed in particle detectors. Rather one believes that, being mini BHs extremely hot, $T_\mathrm{H}>100$ GeV, the emitted energy  can trigger the formation of a photo- and a chromo-sphere, \textit{i.e.}, an electron-positron-photon plasma and a quark-gluon plasma respectively. Such particle atmospheres might result from pair production and bremsstrahlung mechanisms. Accordingly the realistic situation in a particle detector is rather complex: colliding partons are followed by a multiplicity of
particles in the case of BH formation. Both QED and QCD drive such process with different critical temperatures,  namely $T^\mathrm{QED}_\mathrm{c}\sim 50$ GeV and $T^\mathrm{QCD}_\mathrm{c}\sim 175$ MeV respectively. The discrepancy between the two temperatures explains why the  actual BH emission is dominated by hadrons, which result from parton fragmentation. Specifically, one can estimate that the secondary emission consists of 60\% quarks, 15\% gluons, 10\% leptons, 6\% weak bosons, 5\% neutrinos, 1\% photons and smaller fractions of invisible neutrinos, gravitons as well as new particles around $100$ GeV (\cite{CaS06}). In the end, the original Hawking spectrum becomes an effective black body spectrum with a temperature lower than $T_\mathrm{H}$. This is simply due to the energy conservation and the mechanisms of particle proliferation that decrease the average energy per particle.

Additional BH signatures consist in events with a reduced visible energy (due to non-detectable gravitational degrees of freedom emitted in the bulk), with exotic particle production, \textit{e.g.} gluino, squark (\cite{CCN04}), and deformed hadron spectra at high transverse momentum (\cite{ENS09}). In case of remnant formation, the BH event would be indirectly recognized in terms of a significant decrease of the total transverse momentum due to the absence of final decay particles (\cite{KBH05}). Alternatively charged BH remnants might be directly detectable by ionization tracks in time projection chambers.

\section{Update on experimental constraints}
\begin{figure}
\includegraphics[width=0.475\textwidth]{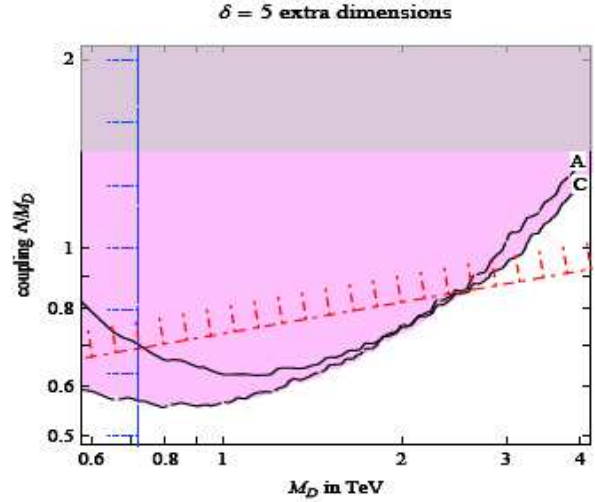}

\caption{(\cite{FGG11}). The shaded area is the bound from virtual graviton exchange at CMS (continuous
line denoted as `C', data after 36/pb), ATLAS (long-dashed line denoted as `A', data after
36/pb). Vertical blue line: bound from graviton emission in (\cite{GiS03}).
Red line: Naive Dimensional Analysis (NDA) estimate of LEP bound from loop graviton exchange.
Upper shading: NDA estimate of the non-perturbative region.
}
\label{figure2}
\end{figure}
\begin{figure}
\includegraphics[width=0.475\textwidth]{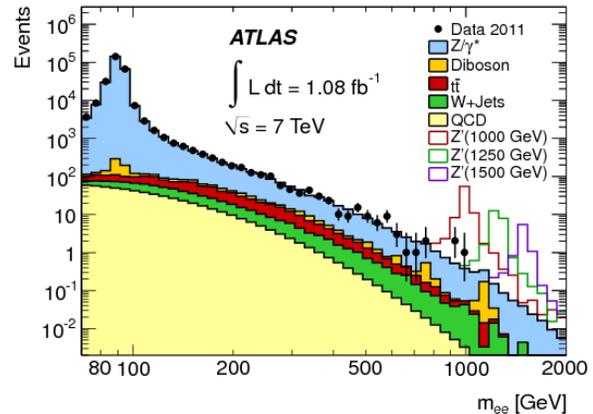}
\caption{(\cite{ATL11}). Dielectron invariant mass ($m_{ee}$) distribution after final selection, compared to the stacked sum of all expected backgrounds, with three example $Z^\prime_\mathrm{SSM}$
signals overlaid. The bin width is constant in $\log m_{ee}$.
}
\label{figure2bis}
\end{figure}
\begin{figure}
\includegraphics[width=0.41\textwidth]{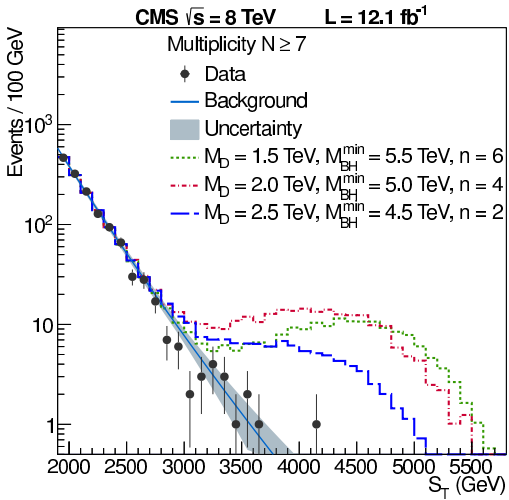}
\caption{(\cite{CMS13}).  Distribution of the total transverse energy, $S_\mathrm{T}$, for events with multiplicity $N\geq 7$ particles in the final state.  Expected semiclassical BH signals are obtained with the BLACK MAX nonrotating BH model (\cite{DSS08}). Here, $M^\mathrm{min}_\mathrm{BH}$ is the minimum BH mass, $M_D$ is the fundamental mass in $D$ dimensions, and $n$ is the number of extra dimensions.  Bottom: The 95\% confidence level (CL) lower limits on the semiclassical BH as a function of $M_D$, for various models. The areas below each curve are excluded by this search. The analysis is performed with CHARYBDIS BH models with or without the stable remnant.}
\label{figure3}
\end{figure}
\begin{figure}
\includegraphics[width=0.41\textwidth]{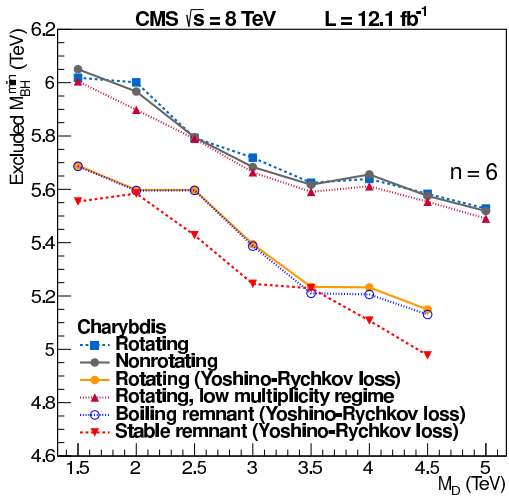}
\caption{(\cite{CMS13}).  The 95\% confidence level (CL) lower limits on the semiclassical BH mass as a function of $M_D$, the fundamental mass in $D$ dimensions for various models. Here $n$ is the number of extra dimensions. The areas below each curve are excluded by this search. The analysis is performed with CHARYBDIS BH models with or without the stable remnant.}
\label{figure4}
\end{figure}

Experimental investigations at the LHC aim to set indirect and direct constraints on relevant parameters, \textit{e.g.} $M_\mathrm{F}$, from events signaling the presence of quantum gravity effects. For instance indirect constraints can be obtained from the exchange of gravitons at tree and one-loop level. Direct constraints are obtained by the observation of BHs.

Franceschini et al. (2011) provided a detailed analysis of processes involving graviton exchange at the light of the latest ATLAS and CMS data. Fig. \ref{figure2} shows the parameter space for the case of $5$ extra-dimension. The function $\Lambda/M_D$ is plotted versus $M_D$, where $\Lambda$ is an ultraviolet cut off and $M_D$ denotes the fundamental mass. Physically $\Lambda$ represents the mass of new states associated to any ultraviolet completion of gravity. Despite $\Lambda$ is an unknown parameter, the advent of LHC has further constrained the parameter space with respect to previous analyses based on the LEP (\cite{GiS03}), making less realistic the occurrence of quantum gravity phenomena at the terascale.

Also the RS models can be tested. In such a set up the KK modes are
not invisible but they should  show up only as spin-2 resonances. As a result if one considers proton collisions in the $e^+e^-$ channel, deviations from conventional Standard Model results in terms of narrow high mass resonances  are expected (\cite{DHR00}). Fig. \ref{figure2bis}, provided by ATLAS Collaboration (2011), shows that this is not the case: the observed invariant mass spectra are consistent with Standard Model expectations.

Finally direct searches of BHs are performed by analyzing possible high transverse energy, high multiplicity events. Also in such a case deviations from the Standard Model are not observed (see Fig. \ref{figure3}). Depending on the model under consideration, analyses from CMS Collaboration (2012, 2013) and ATLAS Collaboration (2014) for collisions at $8$ TeV exclude  BHs with masses  below $4.3-6.2$ TeV (see Fig. \ref{figure4}) in all cases, \textit{i.e.}, non-rotating, rotating BHs, with or without remnant.

\section{Conclusions}

Experimental results are negative. No BHs, no gravitons as well as no quantum gravity effects have been observed at the LHC for collision energy up to $8$ TeV.

The interpretation of these findings requires a careful analysis. A possible explanation for the absence of BHs and any quantum gravity phenomenology at the terascale might be simply due to the fact that the fundamental mass exceeds the LHC design energy, \textit{i.e.}, $M_\mathrm{F}>14$ TeV. If this were the case, we should rely on a ``natural collider", \textit{i.e.}, ultra high energy cosmic rays (UHECRs), whose observed energy $E_\mathrm{CR}$ can reach $10^{20}$ eV, corresponding to $\sim 500$ TeV in the centre of mass frame. The problem with UHECRs is, however, related to meager fluxes. For instance at $E_\mathrm{CR}\sim 10^{19}$, the number of observable events is less that $1$ per kilometer squared, per year. The paucity of events requires large detectors, as it happens for the Pierre Auger Observatory (PAO), which has $1600$ water Cherenkov detectors  distributed over $3000$ square kilometers. This means that, if BHs form from the collision of UHECRs with the upper layers of Earth atmosphere, the PAO can detect  $\sim 100$ BHs in three years (\cite{FeS02}). The current non-observations of events is used to further constraint the value of $M_\mathrm{F}$.

Other competitive bounds can be derived by the conjectured BH production in the scattering of ultrahigh energy cosmic neutrinos on nucleons in the ice or water at AMANDA/IceCube, ANTARES  neutrino telescopes (\cite{KRT02}). This kind of analyses turn to be very sensitive since they can provide additional information about BH branching ratios and the various angle and energy distributions (\cite{AFH02}).

The formation of BHs is certainly a leading test for terascale quantum gravity but only as a sufficient condition. The necessary condition might not be valid. In some energy regimes, terascale quantum gravity could occur without BH formation.  If we consider spacetime geometries admitting  horizon extremization and remnant formation in the Schwarzschild phase, one finds that the minimum energy for BH formation is  $M^\mathrm{min}_\mathrm{BH}\equiv M_\mathrm{remn.}\sim 10$ PeV for $d=5$ and $M_\mathrm{F}=1$ TeV (\cite{Riz06}; \cite{Nic08}; \cite{Gin10}). This is also supported by a recently proposed ghost free, singularity free higher derivative theory of gravity (\cite{BGK12}).  Again the horizon extremization, by deforming Hawking spectra, might lead to quite different BH signatures like a milder emission mostly on the brane (Koch et al. 2005; \cite{CaN08}; \cite{Gin10}; \cite{NiW11}). Other unconventional signatures might arise from the non-thermal decay of BHs (\cite{CCM12}; \cite{ACD13}; \cite{ACC14}). There exist also limitations to the BH production which are rather generic and independent of a specific model: It has been shown that the BH formation might be simply suppressed in order to respect well-established properties of particles, like the lifetime of the proton decay (\cite{SSD06}).

In conclusion, BHs might  be extremely hard to  produce and to detect. Our understanding of several aspects of terascale phenomenology is far from being complete. We recall that the majority of the literature is focused on the Schwarzschild phase, a minor part on the spin-down phase, while little is know about the balding phase.
In addition there exists a variety of effects that are often underestimated, such as the role of the brane tension (\cite{KaK06}), the role of color fields (\cite{MaW00}), the validity of the approximation of quasi-stationary decay.
Finally, the Planck phase is plagued by tremendous difficulties since it connects mini BHs to the problem of the formulation of a quantum theory of gravity. An observation of the Hawking temperature profile in the Planck phase or the formation of remnants could, however, provide crucial indications about the validity of current quantum gravity proposals.

\acknowledgements
This work has been supported by the grant NI 1282/3-1 of the project ``Evaporation of microscopic black holes'' of the German Research Foundation (DFG), by the Helmholtz International Center for FAIR within the framework of the LOEWE program (Landesoffensive zur Entwicklung Wissenschaftlich-\"{O}konomischer Exzellenz) launched by the State of Hesse and partially by the European Cooperation in Science and Technology (COST) action MP0905 ``Black Holes in a Violent Universe''.
The authors are grateful to A. Mazumdar, O. Micu, J. Mureika and D. Stojkovic for valuable comments and references.

\end{document}